\documentclass[cameraready]{Interspeech}

\usepackage{booktabs}
\usepackage{graphicx}
\usepackage{amsmath}
\usepackage{multirow}
\usepackage{siunitx}
\usepackage{tabularx}
\usepackage{makecell, threeparttable} % 表头换行更好看
\usepackage{dblfloatfix} % 改善双栏通栏浮动体放置
\usepackage{adjustbox} % 自动调整宽度
\usepackage[table]{xcolor}

\title{Inside the Latent Flow: Causal Deciphering of Attention Dynamics\\
in Audio Separation Foundation Models}

\author[affiliation={1}, orcid=0009-0009-0389-8307, equalcontribution, correspondingauthor]{Yuxuan}{Chen}
\author[affiliation={2}, orcid=0009-0007-9908-0504, equalcontribution]{Haoyuan}{Yu}
\author[affiliation={3}, orcid=0009-0008-3693-5576, equalcontribution]{Peize}{He}

\address{
    $^1$ Jilin University,
    $^2$ Hunan University, \\
    $^3$ University of Electronic Science and Technology of China 
}

\email{yxchen5522@mails.jlu.edu.cn, y15352176976@hnu.edu.cn, 2023300904027@std.uestc.edu.cn}

\keywords{audio separation, mechanistic interpretability, diffusion transformers, flow matching}

\usepackage{comment}

\begin{document}
\maketitle

%==================================================================================
% Content
\begin{abstract}
Flow-matching transformers achieve strong audio separation, yet their attention dynamics are opaque. We adapt established causal-intervention principles into a deterministic, inference-time probing protocol for SAM Audio. Orthogonal probing uncovers a dual-pathway text-conditioning mechanism: additive injections control semantic identity, while cross-attention refines acoustic structure. We observe an asynchronous layerwise convergence: stable layers build temporal scaffolds early, whereas fast layers continue resolving artifacts during sampling. The model also attenuates temporal segmentation cues to maintain continuous-flow stability. Using these insights, we propose \textbf{L}ayer-\textbf{S}elective \textbf{A}ttention \textbf{C}aching (\textbf{LSAC}), a training-free acceleration method that caches attention in stable layers. Across acoustic complexities, LSAC cuts self-attention computation by ${\sim}25\%$ with negligible quality loss and yields up to 6.7x higher quality retention than naive step reduction.
\end{abstract}

%%%%%%%%%%%%%%%%%%%%%%%%   正文开始   %%%%%%%%%%%%%%%%%%%%

\section{Introduction}
\label{sec:intro}

The promptable segmentation paradigm \cite{kirillov2023segmentanything} has successfully transitioned from computer vision to audition, establishing a robust baseline for universal sound separation \cite{shi2025samaudio}. Modern audio foundation models integrate continuous flow matching \cite{lipman2023flowmatching} with diffusion transformers \cite{peebles2023dit} to process diverse multimodal conditions \cite{evans2024stableaudioopen, liu2024audioldm2}. These architectures isolate overlapping sound sources by integrating ordinary differential equations over continuous noise distributions within a compressed latent space \cite{ho2020ddpm}. Recent extensions encompass language-queried rectified flow \cite{yuan2024flowsep}, multi-granularity latent diffusion \cite{chae2025mgeldm}, and attention-based acceleration \cite{tan2024litefocus, chen2025tfattn, bellur2024biomimetic}. Despite strong separation fidelity, the internal mechanisms governing multimodal feature injection and spatiotemporal routing during integration remain poorly understood.

Current explainability practices in generative modeling rely predominantly on heuristics derived from computer vision. In visual diffusion models, cross-attention maps are universally treated as precise spatial grounding mechanisms, utilized for token-to-pixel attribution \cite{tang2023daam}, masked diffusion guidance \cite{endo2024mag}, and semantic image editing \cite{alaluf2023attendexcite}. The audio community frequently assumes these spatial alignment principles transfer directly to the temporal flow of acoustic generation \cite{rombach2022ldm, hertz2023prompt2prompt}. However, recent analyses of diffusion transformers for audio denoising \cite{li2024ditaudio} and causal deciphering via diffusion \cite{duan2024causal} suggest that directly transferring these assumptions may be misleading, and it remains unclear whether cross-attention truly functions as a temporal anchor in audio architectures.

We identify a critical methodological gap: attention is not necessarily explanation \cite{jain2019attentionnotexplanation}, and passively observing attention distributions cannot decouple the entangled nonlinear interactions in diffusion models \cite{liu2024understanding_sd_attention, chefer2021beyondattention}. Achieving mechanistic interpretability demands causal interventions that isolate and quantify specific architectural dependencies.

To bridge this gap, we adapt causal-intervention principles \cite{pearl2009causality} into a deterministic, inference-time probing framework. This approach manipulates intermediate representations during the ordinary differential equation trajectory without altering pretrained weights. We apply this framework to the SAM Audio~\cite{shi2025samaudio} to decipher its latent dynamics.

In summary, we make the following contributions:

\begin{itemize}[leftmargin=*, nosep]
    \item \textbf{Mechanistic Discovery.} Through orthogonal probing, we establish an asymmetric dual pathway conditioning mechanism: additive injection drives semantic identity while cross-attention governs acoustic structure. This finding challenges the prevailing assumption that cross-attention serves as the primary semantic grounding mechanism in audio diffusion models.
    
    \item \textbf{Causal Verification.} We demonstrate an asynchronous scaffold and sculpt convergence dynamic across self-attention layers, providing direct causal evidence that layer-wise attention maturation follows a non-uniform temporal schedule during the ODE trajectory.
    
    \item \textbf{Phenomenon Characterization.} We identify a prior suppression phenomenon wherein the network actively hibernates temporal span capabilities to preserve continuous flow trajectory stability, revealing an implicit trade-off between representational capacity and generative coherence.
    
    \item \textbf{Practical Method.} Leveraging these insights, we propose \textbf{L}ayer-\textbf{S}elective \textbf{A}ttention \textbf{C}aching (\textbf{LSAC}) , a training-free acceleration strategy that exploits asynchronous convergence to strictly dominate naive step reduction across diverse acoustic complexities. LSAC scales efficiently to 3B parameter models while maintaining separation fidelity.
\end{itemize}

\begin{figure}[t]
  \centering
  \includegraphics[width=0.9\linewidth]{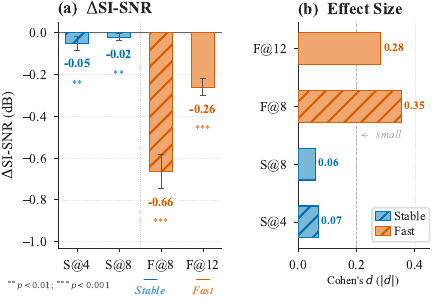}
  \vspace{-10pt}
  \captionsetup{font=footnotesize} % 或 font=foot
  \caption{Acoustic impact under layer specific causal freezing (Clean Tier, $N{=}200$ paired runs per layer). Freezing Stable layers midway yields practically negligible degradation, whereas interrupting Fast layers halts fine grained artifact reduction.}
  \label{fig:scaffold_sculpt}
  \vspace{-4pt}
\end{figure}

%%%%%%%%%%%%%%%%%%%%%%%%%%%%   Method   %%%%%%%%%%%%%%%%%%%%%%%%%%%%%%%%%%%%%
\section{Method}
\label{sec:method}

We define a framework of deterministic inference time interventions \cite{pearl2009causality} to probe the underlying mechanisms of the audio diffusion transformer. These operations manipulate the continuous probability flow trajectory \cite{song2021scorebasedsde} during inference without altering the pretrained network weights.

\subsection{Orthogonal Probing}

Orthogonal probing isolates the two text conditioning pathways. The diffusion transformer layer processes text conditions $\mathbf{c}$ via an additive residual injection \cite{perez2018film} and a standard cross attention mechanism \cite{vaswani2017attention}. We formulate the additive modulation $\mathbf{m}$ combining the projected text and the time embedding $\mathbf{t}_{emb}$ as follows:
\begin{equation}
    \mathbf{m} = \text{proj}(\mathbf{c}) + \mathbf{t}_{emb}
\end{equation}

The cross attention operation over the latent query $\mathbf{Q}$, text key $\mathbf{K}_c$, and text value $\mathbf{V}_c$ is defined as:
\begin{equation}
    \text{CA}(\mathbf{Q}, \mathbf{K}_c, \mathbf{V}_c) = \text{softmax} \left( \frac{\mathbf{Q}\mathbf{K}_c^\top}{\sqrt{d}} \right) \mathbf{V}_c
\end{equation}

We define three ablation conditions: \textit{Zeroed CA} ($\text{CA}\to\mathbf{0}$), \textit{Additive Zeroed} ($\text{proj}(\mathbf{c})\to\mathbf{0}$), and \textit{Forced Uniform} (softmax weights $\to 1/N_{\text{text}}$), and measure the impact using physically orthogonal metrics.

\subsection{Causal Freezing}

Causal freezing localizes the integration step at which physical mappings stabilize. Let $\mathbf{A}^{(l)}(t)$ denote the self attention matrix of layer $l$ at integration step $t$ \cite{abnar2020attentionflow}. We classify layers by entropic rate of change. The attention entropy $H^{(l)}(t) = -\sum_j A^{(l)}_{ij} \log A^{(l)}_{ij}$ (averaged over all heads and query positions) yields a mean absolute change $\Delta E^{(l)} = \frac{1}{T-1} \sum_{t=1}^{T-1} |H^{(l)}(t{+}1) - H^{(l)}(t)|$. Layers with $\Delta E^{(l)} \leq \theta$ (median threshold across all layers) are classified as \textit{stable}; others as \textit{fast}. The freeze step $\tau$---swept here during analysis, but fixed by LSAC in Eq.~\ref{eq:threshold-configurations}---sets the step from which we clamp attention:
\begin{equation}
    \mathbf{A}^{(l)}(t) \leftarrow \mathbf{A}^{(l)}(\tau), \quad \forall\, t > \tau
\end{equation}

This intervention quantifies the acoustic contribution of the interrupted integration steps.

\subsection{Gate Hijacking}

Gate hijacking probes the internal geometric capacity for temporal segmentation. The default fusion equation for span prompts utilizes a learnable gating parameter $\gamma$. The latent hidden state $\mathbf{h}$ is updated using the temporal span embedding $\mathbf{e}_{span}$ via the following modulation:
\begin{equation}
    \mathbf{h} \leftarrow \mathbf{h} + \tanh(\gamma) \mathbf{e}_{span}
\end{equation}

The pretrained parameter defaults to a negative scalar $\gamma = -0.14$. We hijack this mechanism by forcibly setting $\gamma = +5.0$. We include a shuffled temporal alignment control group alongside the hijacked gate to rigorously verify boundary specific topological capabilities. We quantify topological selectivity via the head-averaged Block Ratio $\text{BR} = {\sum_{ij} A_{ij} M_{ij}}/{(\sum_{ij} A_{ij}(1{-}M_{ij}) + \epsilon)}$, where $\mathbf{M}$ is a binary span mask and $\epsilon = 10^{-8}$ ensures numerical stability.

\subsection{Layer Selective Attention Caching}

Layer Selective Attention Caching translates causal freezing into an engineering strategy. For a designated subset of layers, if the current step $t$ exceeds that layer's freeze step $\tau$ (Eq.~\ref{eq:threshold-configurations}), the model skips the computationally expensive $\mathcal{O}(T^2 d)$ query-key multiplications to reuse the cached $\mathbf{A}^{(l)}(\tau)$;
the value matrix $\mathbf{V}$ is recomputed each step since $\mathbf{A}$ converges earlier than $\mathbf{V}$ in stable layers. Self-attention constitutes 29\% of total inference FLOPs.\footnote{For SAM-Audio-Small with a 16-step Euler solver, the baseline requires 211.4 GFLOPs, of which 61.4 GFLOPs are self-attention.}

Let $T$ equal the total number of integration steps. We define three distinct threshold configurations for the stable layers and the moderate convergence layers:
\begin{equation}
\label{eq:threshold-configurations}
\begin{aligned}
(\tau_{\mathrm{stable}}, \tau_{\mathrm{moderate}})
&=
\begin{cases}
\bigl(\frac{T}{2},\, T\bigr), 
    & \textsc{Safe}, \\
\bigl(\frac{3T}{8},\, \frac{5T}{8}\bigr), 
    & \textsc{Balanced}, \\
\bigl(\frac{T}{4},\, \frac{T}{2}\bigr), 
    & \textsc{Aggressive}.
\end{cases}
\end{aligned}
\end{equation}

A compute-matched naive truncation uniformly truncates total integration steps without selective caching.

\begin{figure*}[!t]
  \centering
  \includegraphics[width=\textwidth]{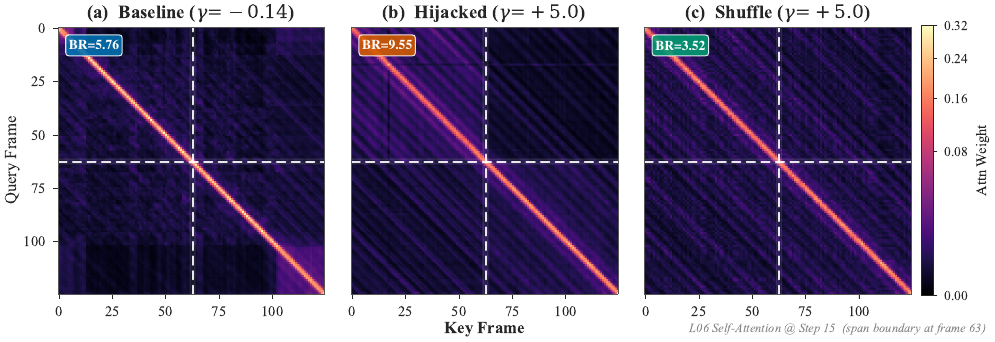}
  \vspace{-5pt}
  \caption{Visual proof of dormant geometry (L06 self-attention, integration step 15). Forcing a positive gate activates latent block diagonal capacities strictly aligned with boundaries. The model actively suppresses these features during standard optimization.}
  \label{fig:gate_hijack}
\end{figure*}

%%%%%%%%%%%%%%%%%   实验设置   %%%%%%%%%%%%%%%%%%%%%%%%%%
\section{Experimental Setup}
\label{sec:setup}

We conduct causal probing on the open source SAM Audio Small model. This model features 12 transformer layers and utilizes a 16 step Euler solver. It processes audio inside a 25 Hz discrete autoencoder latent space \cite{kumar2023dac}. To eliminate capacity induced behavioral discrepancies, we execute parallel random sampling on the 3 billion parameter large variant containing 22 layers. The observed causal rules and acceleration ratios remain consistent across the two model capacities we evaluate.

The evaluation set spans three complexity tiers and includes over 10000 independent ordinary differential equation runs. The Clean tier comprises pure speech mixtures from LibriSpeech \cite{panayotov2015librispeech} emphasizing challenging cross gender overlap. The Noisy tier injects 5 dB of steady state white noise. The Env tier contains cross domain audio sourced from ESC-50 \cite{piczak2015esc50} and FSD50K \cite{fonseca2020fsd50k}.

We establish a physically orthogonal evaluation scheme. We measure the acoustic axis with Scale-Invariant Signal-to-Noise Ratio \cite{luo2019convtasnet} and Signal-to-Artifacts Ratio \cite{vincent2006bsseval}, and the semantic axis with Short-Time Objective Intelligibility \cite{taal2011stoi} and Perceptual Evaluation of Speech Quality \cite{itu2001pesq}: a direct semantic, not purely energy-based, evaluation of the conditioning pathways. Zero shot blind separation exhibits high intrinsic variance with a standard deviation near 16.8 dB. We strictly utilize paired t-tests and report Cohen's d \cite{cohen1988power} to quantify standardized effect sizes. The effect sizes are categorized as negligible, small, medium, or large. A Bonferroni correction is applied for all multiple comparisons.

%%%%%%%%%%%%%%%%%%%%%%%%%%%%%%   实验结果 Results   %%%%%%%%%%%%%%%%%%%%%%%%%%%%%%%%%%%%
\section{Results}
\label{sec:results}

\begin{table}[t]
\centering
\captionsetup{font=footnotesize}
\caption{Orthogonal probing of conditioning pathways (Small model, Clean Tier; $N{=}2500$ paired runs): $\Delta$ vs.\ Baseline, Cohen's $d$ in parentheses. All effects significant under Bonferroni-corrected paired $t$-tests ($p{<}0.001$; $\alpha{=}0.0042$).}
\label{tab:orthogonal}

\footnotesize
% ---- local palette (group-scoped; no \definecolor) ----
\def\tbAcc{\color[RGB]{52,77,105}}    % muted slate accent
\def\tbMut{\color[RGB]{122,122,122}}  % quiet gray
% ---- raw cell content (used for probing and inside boxes) ----
\def\tbVraw#1#2{$#1$\,{\tiny\tbMut($#2$)}}                 % value (d)
\def\tbBraw#1#2{{\tbAcc$\mathbf{#1}$}\,{\tiny\tbMut($#2$)}}% headline col
% ---- geometry ----
\setlength{\tabcolsep}{2.5pt}
\renewcommand{\arraystretch}{1.3}
\setlength{\aboverulesep}{0.12ex}
\setlength{\belowrulesep}{0.32ex}
% ---- adaptive column width: max(distributed, widest content + pad) -
% register-free: maxima live in macros, immune to kernel scratch use.
\def\tbMax{0pt}%
\def\tbProbe#1{\sbox0{#1}\ifdim\wd0>\tbMax\relax\edef\tbMax{\the\wd0}\fi}%
\tbProbe{\tbVraw{-0.101}{-0.44}}\tbProbe{\tbVraw{-0.206}{-0.82}}%
\tbProbe{\tbBraw{-14.13}{-0.74}}\tbProbe{\tbBraw{-10.99}{-0.83}}%
\tbProbe{\textbf{Uniform CA}}\tbProbe{\textbf{Zeroed CA}}%
\edef\tbMax{\the\dimexpr\tbMax+2pt\relax}% breathing room per column
\sbox0{$\Delta$\textbf{SI-SNR}}\edef\tbLbl{\the\wd0}% widest row label
\edef\tbDist{\the\dimexpr(\linewidth-\tbLbl-\tabcolsep*10)/4\relax}%
\ifdim\tbDist>\tbMax\relax\edef\tbW{\tbDist}\else\edef\tbW{\tbMax}\fi
% ---- boxed cells: centering guaranteed by \makebox, not by glue ----
\def\tbV#1#2{\makebox[\tbW][c]{\tbVraw{#1}{#2}}}
\def\tbB#1#2{\makebox[\tbW][c]{\tbBraw{#1}{#2}}}
\def\tbD{\makebox[\tbW][c]{\tbMut---}}
\def\tbHone#1{\makebox[\tbW][c]{#1}}                   % 1-line header
\def\tbHtwo#1#2{\makebox[\tbW][c]{\begin{tabular}[c]{@{}c@{}}#1\\#2\end{tabular}}}% 2-line, midpoint on baseline
\def\tbM#1{$\Delta$\textbf{#1}}                        % metric label
\resizebox{\linewidth}{!}{%
\begin{tabular}{l c c c >{\columncolor[RGB]{237,241,246}}c}
\toprule
\rowcolor[RGB]{245,245,243}
\textbf{Metric} & \tbHone{\textbf{Baseline}} & \tbHtwo{\textbf{Forced}}{\textbf{Uniform CA}}
               & \tbHone{\textbf{Zeroed CA}} & \tbHtwo{\textbf{Additive}}{\textbf{Zeroed}} \\
\arrayrulecolor[RGB]{110,110,110}\midrule
\tbM{SI-SNR} & \tbD & \tbV{-4.44}{-0.29}  & \tbV{-7.99}{-0.49}  & \tbB{-14.13}{-0.74} \\
\tbM{STOI}   & \tbD & \tbV{-0.101}{-0.44} & \tbV{-0.206}{-0.82} & \tbB{-0.219}{-0.89} \\
\tbM{PESQ}   & \tbD & \tbV{-0.12}{-0.35}  & \tbV{-0.17}{-0.38}  & \tbB{-0.18}{-0.37}  \\
\tbM{SAR}    & \tbD & \tbV{-3.40}{-0.29}  & \tbV{-9.85}{-0.82}  & \tbB{-10.99}{-0.83} \\
\arrayrulecolor{black}\bottomrule
\end{tabular}}%
\end{table}

%%%   4.1节 双路条件   %%%
\subsection{Dual Pathway Conditioning}
\label{ssec:dual_pathway}

We observe an asymmetric division of labor within the text conditioning mechanisms. Table \ref{tab:orthogonal} details the orthogonal physical responses to conditioning ablation. 

The additive injection pathway dominates semantic identity. Ablating the additive vector yields the largest degradation on the semantic axis (STOI: $\Delta = -0.219$, $d = -0.89$, $p < 0.001$). This effect intensifies on the 3B variant, where the Environmental tier exhibits $\Delta$STOI~$= -0.336$ ($d = -1.14$), confirming that additive dominance persists at the 3B scale. Intuitively, the additive vector functions as a macroscopic steering mechanism. It globally modulates the scale and shift of every latent token to construct the macroscopic feature manifold required for recognizing the semantic target class. The model loses fundamental structural knowledge regarding the target sound when this pathway is disconnected.

Conversely, zeroing the cross-attention pathway induces a massive degradation on the
acoustic axis (SAR $= -9.85$~dB, $d = -0.82$, $p < 0.001$). This confirms that cross attention is primarily responsible for rendering high frequency acoustic textures and source separation sharpness. Rather than providing rigid temporal grounding, the cross attention matrix acts as a localized microscopic modulator. It aligns transient phase structures without dictating the broader semantic trajectory.

The Forced Uniform condition, which replaces learned attention fluctuations with a flat residual, produces comparatively mild degradation (STOI $d = -0.44$, SAR $d = -0.29$), indicating that even crude alignment suffices when the additive pathway remains intact.

\subsection{Asynchronous Scaffold and Sculpt Dynamics}
\label{ssec:scaffold}

The internal layers exhibit distinct and asynchronous convergence phases during ordinary differential equation integration. Figure \ref{fig:scaffold_sculpt} illustrates the acoustic impact of layer specific causal freezing. 

Stable layers (L1, L6, L9) can be safely frozen as early as Step 4. The resulting degradation in SI-SNR is 0.05 dB, corresponding to a negligible effect size ($d = 0.07$). This proves that macroscopic temporal smoothing is fully constructed in the earliest phases of the trajectory. These layers function exclusively as scaffolders that establish the low frequency envelope and reject initial interference before converging into a static structure.

Fast layers (L0, L2, L3, L8, L10) demonstrate continuous activity. Freezing these layers at Step 8 triggers a degradation of 0.66 dB with a small effect size ($d = 0.35$, $p < 10^{-35}$). This statistically significant drop confirms that fast layers remain actively engaged in resolving fine grained artifacts during the final integration steps. 

Delaying the freeze to Step 12 limits the degradation to 0.26 dB ($d = 0.28$), confirming that high frequency transients only become resolvable as the latent state approaches the clean distribution near the trajectory endpoint. Premature freezing at Step 8 thus preserves the temporal scaffold but leaves fine acoustic details unresolved.

\subsection{Prior Suppression of Temporal Geometry}
\label{ssec:gate_hijack}

The foundation model natively possesses the geometric capability to segment discrete boundaries. However, it actively suppresses this function during inference to favor continuous flow. Figure~\ref{fig:gate_hijack} visualizes the self-attention matrices under gate hijacking.

Under the default pretrained gating parameter ($\gamma = -0.14$), the model largely ignores discrete temporal span boundaries. The self-attention matrix retains a diffuse diagonal structure representing localized temporal smoothing, with a Block Ratio of 5.76. When we forcibly hijack the gate to $+5.0$, the L06 layer Block Ratio surges by 66\% to 9.55. The heatmap immediately reveals a perfect block diagonal topology aligned strictly with the span boundary. This forced geometric activation simultaneously causes the SI-SNR to collapse by 14.6\,dB.

To verify causality, we apply a shuffled temporal alignment alongside the $+5.0$ gate. The topological structure instantly plummets by 63\% to a Block Ratio of 3.52. This boundary-specificity is layer-dependent: L01 shows no response (BR~$2.2{\to}2.1$), whereas L09 exhibits a 150\% surge, suggesting that geometric capacity concentrates in mid-to-late layers. Across a 15-point $\gamma$ sweep ($\gamma \in [-1, 6]$, $N{=}3{,}400$ paired runs), the pooled Spearman rank correlation between BR and SI-SNR yields $\rho = 0.607$ ($p = 3.8 \times 10^{-4}$), providing strong quantitative evidence that the block diagonal phenomenon is an inherent boundary-specific capacity rather than a numerical perturbation artifact.

We interpret this suppression as a consequence of the generative training objective: continuous flow matching favors smooth probability-density transport, whereas discrete span prompts impose hard step functions that conflict with this manifold. Consistent with this view, the model converges to a negative gate ($\gamma = -0.14$), and the strong BR--SI-SNR coupling ($\rho = 0.607$) suggests that forcing the discrete prior actively degrades separation. This may reflect a learned strategy to suppress a rigid prior in favor of a self-organized separation path. A full mechanistic account of how this gate is learned is left to future work.

% 4.4节 
\subsection{Complexity-Aware Inference Acceleration}
\label{ssec:lsac}

\begin{figure}[t]
  \centering
  \includegraphics[width=\linewidth]{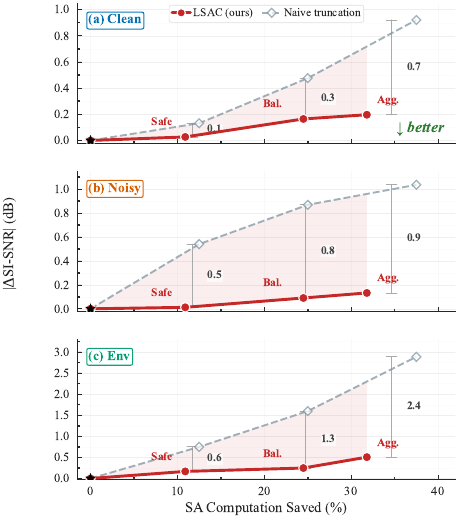}
  \vspace{-5pt}
  \caption{Pareto efficiency of LSAC. The strategy strictly dominates naive baseline truncation across all acoustic complexities without requiring retraining.}
  \label{fig:pareto_frontier}
\end{figure}

Building on the asynchronous convergence identified in Section~\ref{ssec:scaffold}, we introduce \textbf{L}ayer-\textbf{S}elective \textbf{A}ttention \textbf{C}aching (\textbf{LSAC}). Unlike global feature-caching methods that skip entire transformer blocks \cite{ma2023deepcache} or frequency-domain caching \cite{tan2024litefocus}, LSAC selectively targets the attention topology based on layer-specific convergence dynamics. Since stable layers complete temporal scaffolding early, their attention matrices can be safely cached for subsequent integration steps. Figure~\ref{fig:pareto_frontier} confirms that this approach achieves strict Pareto dominance over naive step reduction across all three complexity tiers.

% =====================================================================
%  Cross-tier validation table — full-band elegant version (FIXED)
%  REQUIRED PACKAGES (preamble): booktabs, multirow, graphicx
%      (already needed for your figures) and
%      \usepackage[table]{xcolor}
%      (if xcolor is already loaded by the template, put
%       \PassOptionsToPackage{table}{xcolor} BEFORE \documentclass)
%
%  COLLISION-PROOF: helper macros use \def (locally shadows any
%  same-named command from your preamble, e.g. soul's \hl situation,
%  and auto-restores after \end{table}) with rare tb-prefixed names.
%  All colors inline [RGB]; nothing leaks into global scope.
% =====================================================================
\begin{table}[t]
\centering
\captionsetup{font=footnotesize}
\caption{Cross-tier quality validation (Small model): $|\Delta\mathrm{SI\mbox{-}SNR}|$ (dB) vs.\ the full 16-step Euler baseline; $N{=}200$ paired runs/cell.}
\label{tab:auxiliary}
\footnotesize
% ---- local palette (group-scoped; no \definecolor) ----
\def\tbAcc{\color[RGB]{52,77,105}}    % muted slate accent
\def\tbMut{\color[RGB]{122,122,122}}  % quiet gray for stats
% ---- local cell macros (\def = safe even if names exist globally) ----
\def\tbPar#1{\,{\tiny\tbMut(#1)}}                       % Cohen's d
\def\tbVal#1#2{\leavevmode$#1$\tbPar{#2}}               % value
\def\tbBest#1#2{\leavevmode{\tbAcc$\mathbf{#1}$}\tbPar{#2}}% column best
\def\tbSig#1#2#3{\leavevmode$#1^{\tbMut#2}$\tbPar{#3}}  % value + sig.
\def\tbHdr#1{\textbf{#1}\,{\scriptsize\tbMut$\downarrow$}}
\def\tbChip{{\setlength{\fboxsep}{2pt}\setlength{\fboxrule}{0.5pt}{\tbAcc\fbox{\textbf{LSAC}}}}}
% ---- geometry ----
\setlength{\tabcolsep}{2.5pt}
\renewcommand{\arraystretch}{1.12}
\setlength{\aboverulesep}{0.12ex}
\setlength{\belowrulesep}{0.32ex}
% ---- adaptive column width: max(distributed, widest content) -------
% probes run in YOUR template's fonts; maxima live in MACROS (never in
% low \dimen scratch registers, which the kernel may clobber between
% probe calls); \resizebox is a no-op (scale=1) when the column fits.
\def\tbMax{0pt}%
\def\tbProbe#1{\sbox0{#1}\ifdim\wd0>\tbMax\relax\edef\tbMax{\the\wd0}\fi}%
\tbProbe{\tbSig{2.6}{***}{0.16}}\tbProbe{\tbSig{5.0}{***}{0.31}}%
\tbProbe{\tbSig{7.3}{***}{0.46}}\tbProbe{\tbSig{4.1}{***}{0.14}}%
\edef\tbMax{\the\dimexpr\tbMax+2pt\relax}% breathing room per column
\sbox0{\textbf{DeepCache}~\cite{ma2023deepcache}}\edef\tbLbl{\the\wd0}% widest col-1 item
\edef\tbDist{\the\dimexpr(\linewidth-\tbLbl-1.6cm-\tabcolsep*10)/3\relax}%
\ifdim\tbDist>\tbMax\relax\edef\tbW{\tbDist}\else\edef\tbW{\tbMax}\fi
\resizebox{\linewidth}{!}{%
\begin{tabular}{c >{\centering\arraybackslash\scriptsize\itshape}p{1.6cm} *{3}{>{\centering\arraybackslash}p{\tbW}}}
\toprule
\rowcolor[RGB]{245,245,243}
\multicolumn{2}{c}{\textbf{Condition}} & \tbHdr{Clean} & \tbHdr{Noisy} & \tbHdr{Env} \\
\arrayrulecolor[RGB]{110,110,110}\midrule
\rowcolor[RGB]{237,241,246} & Safe
& \tbBest{0.5}{0.02} & \tbBest{0.1}{0.01} & \tbBest{0.2}{0.01} \\[-1pt]
\rowcolor[RGB]{237,241,246} & Balanced
& \tbVal{0.2}{0.02} & \tbBest{0.1}{0.00} & \tbVal{0.3}{0.01} \\[-1pt]
\rowcolor[RGB]{237,241,246}
\multirow{-3}{*}{\shortstack[c]{\tbChip\\[1.5pt]{\tbAcc\scriptsize\itshape(ours)}}}
& Aggressive
& \tbSig{2.5}{***}{0.08} & \tbVal{0.6}{0.02} & \tbBest{0.2}{0.01} \\
\arrayrulecolor[RGB]{175,175,175}\cmidrule(lr){1-5}
& 14-step
& \tbSig{2.6}{***}{0.16} & \tbSig{1.3}{*}{0.04} & \tbSig{1.3}{**}{0.04} \\[-1pt]
& 12-step
& \tbSig{5.0}{***}{0.31} & \tbSig{3.2}{***}{0.10} & \tbSig{4.1}{***}{0.14} \\[-1pt]
\multirow{-3}{*}{\textbf{Naive}}
& 10-step
& \tbSig{7.3}{***}{0.46} & \tbSig{4.7}{***}{0.16} & \tbSig{7.4}{***}{0.26} \\
\cmidrule(lr){1-5}
& Skip-2
& \tbVal{0.9}{0.02} & \tbVal{0.7}{0.03} & \tbSig{1.1}{*}{0.04} \\[-1pt]
& Skip-3
& \tbVal{0.8}{0.02} & \tbVal{0.6}{0.03} & \tbVal{0.9}{0.03} \\[-1pt]
\multirow{-3}{*}{\textbf{DeepCache}~\cite{ma2023deepcache}}
& Skip-4
& \tbVal{1.2}{0.05} & \tbVal{0.8}{0.01} & \tbSig{1.9}{*}{0.08} \\
\arrayrulecolor{black}\bottomrule
\end{tabular}}\par
\vspace{4pt}
\parbox{\columnwidth}{\scriptsize
Signed SI-SNR is not monotonically related to perceptual quality for generative ODE-based separators; we therefore report deviation magnitude (behavioral fidelity). Parentheses: Cohen's $d$. Bonferroni-corrected paired $t$-tests: $^{***}p<0.001$, $^{**}p<0.01$, $^{*}p<0.05$; unmarked = n.s.}
\vspace{-6pt}
\end{table}

At an equivalent self-attention computation saving of approximately 25\%, the balanced configuration demonstrates strong robustness. In the Clean tier, LSAC yields a \(2.5\times\) advantage over the naive baseline (0.19 dB degradation versus 0.48 dB). In the Noisy tier, the caching strategy incurs a negligible 0.13 dB degradation, whereas the naive baseline degrades by 0.87 dB, resulting in a \(6.7\times\) advantage. The Environmental tier requires the most robust processing due to non-speech spectral complexity; here, caching degrades performance by only 0.30 dB, while the naive baseline exhibits substantial failure (1.60 dB), maintaining a \(5.3\times\) advantage. On the 3B variant (22 layers, 11 stable), LSAC-Balanced achieves \(|\Delta|=0.01\) dB on Clean versus DeepCache-Skip2's 0.37 dB, a \(37\times\) quality advantage, confirming that convergence-guided caching scales to production-sized architectures.

Table~\ref{tab:auxiliary} further confirms that caching preserves cross-tier behavioral fidelity. Extended cross-tier validation with Bonferroni-corrected paired $t$-tests across 4{,}200 ODE runs confirms that LSAC-Safe and LSAC-Balanced show \textit{statistically non-significant} degradation ($p > 0.05$) on both Noisy and Env tiers, whereas all naive truncation conditions exhibit significant degradation. LSAC is orthogonal to solver choice: caching attention \textit{within} each step is complementary to higher-order solvers that reduce step count.

%%%%%%%%%%%%%%%%   引用图三   %%%%%%%%%%%%%%%%%

%%%%%%%%%%%%%%%%%%%%%%%%%%%%%%%%%%%%%%%%%%   结论段   %%%%%%%%%%%%%%%%%%%%%%%%%%%%%%%%%%%%%%
\section{Conclusion}
\label{sec:conclusion}

This work applies strict inference time causal interventions to decipher the latent flow dynamics of audio diffusion foundation models. We quantitatively establish a dual pathway text conditioning mechanism where additive injections govern semantic identity while cross attention resolves acoustic textures. We map abstract integration steps to physical acoustics, identifying an asynchronous scaffold and sculpt phase separation among network layers. We also demonstrate that the model actively suppresses its native temporal segmentation capabilities to maintain continuous manifold flow. Deriving practical value from interpretability, LSAC yields up to $6.7\times$ quality retention advantage over naive step reduction and generalizes across acoustic complexities with statistically non-significant degradation. Future work will explore dynamic caching thresholds based on real time signal to noise estimation. Our study is confined to the SAM Audio family (Small and 3B); whether the dual-pathway division and asynchronous convergence transfer to other conditioning schemes remains open.

%%%%%%%%%%%%%%%%%%%%%%%%%%%%%%   正文结束   %%%%%%%%%%%%%%%%%%%%%%%%%%%%%%%%

\clearpage   % 强制参考文献从下一页开始
\nocite{*}   % 强制把 .bib 里所有条目都当作“被引用”

\begingroup  % 开启分组，分组里的设置在分组结束后会恢复，不影响外部内容
\raggedbottom

\section{Generative AI Use Disclosure}
We used a generative AI tool to assist with language editing and polishing of the manuscript, including improving grammar, clarity, and readability.
The tool was not used to generate scientific content, experimental results, or conclusions.
All coauthors reviewed the final manuscript and take full responsibility for it.
\vspace{-2pt}

\bibliographystyle{IEEEtran}
\bibliography{mybib}
\endgroup   % 结束分组

\end{document}